\documentclass[aps, onecolumn]{revtex4}

\usepackage[dvips]{graphicx}
\usepackage{textcomp}

\begin{document}

\title{Realization of XNOR and NAND spin-wave logic gates}

\author{T. Schneider}\email{tschneider@physik.uni-kl.de}
\author{A.A. Serga}
\author{B. Leven}
\author{B. Hillebrands}
\affiliation{Fachbereich Physik und Forschungsschwerpunkt MINAS,
Technische Universit\"at Kaiserslautern, 67663 Kaiserslautern,
Germany}

\author{R.L. Stamps}
\author{M.P. Kostylev}
\affiliation{School of Physics, University of Western Australia,
Crawley, WA 6009, Australia}

\begin{abstract}
We demonstrate the functionality of spin-wave logic XNOR and NAND
gates based on a Mach-Zehnder type interferometer which has arms
implemented as sections of ferrite film spin-wave waveguides.
Logical input signals are applied to the gates by varying either the
phase or the amplitude of the spin waves in the interferometer arms.
This phase or amplitude variation is produced by Oersted fields of
dc current pulses through conductors placed on the surface of the
magnetic films.
\end{abstract}

\maketitle

Although commonly used for data storage applications, there have
been relatively few attempts to employ magnetic phenomena for
performing logical operations. The suggested concepts include the
control of domain wall movement \cite{1}, of magnetoresistance of
individual magnetic elements \cite{2}, and of a magnetostatic field
of a set of magnetic nanoelements \cite{3}. Yet another concept is
using spin-wave interferometers. It was discussed theoretically in
Refs. \cite{4,5,6}, but there was only one experimental
demonstration of spin wave logic gate functionality \cite{7}, where
an one-input NOT gate was implemented in a interferometer-like
geometry. In the present work we experimentally demonstrate the
functionality of more complicated logic gates based on spin waves.

The fabricated prototype of a XNOR logic gate is a direct extension
of the NOT gate from Ref. \cite{7} which was based on a Mach-Zehnder
interferometer. For its implementation the reference interferometer
arm of the NOT gate is replaced by an arm identical to the signal
arm. Controlling phases accumulated by the spin waves in both arms
allows one to perform the XNOR operation.

Demonstrating the functionality of a NAND logic gate is a
considerable step forward in the development of spin wave logic
compared to the NOT and XNOR gates. Firstly because the NAND
function belongs to a class of universal functions which means that
combining NAND gates allows one to construct gates of other types.
Secondly because for its implementation, we use here a new physical
principle: direct control of spin wave amplitudes in the
interferometer arms.

\begin{figure}[h!tp]
  \centering
  \includegraphics{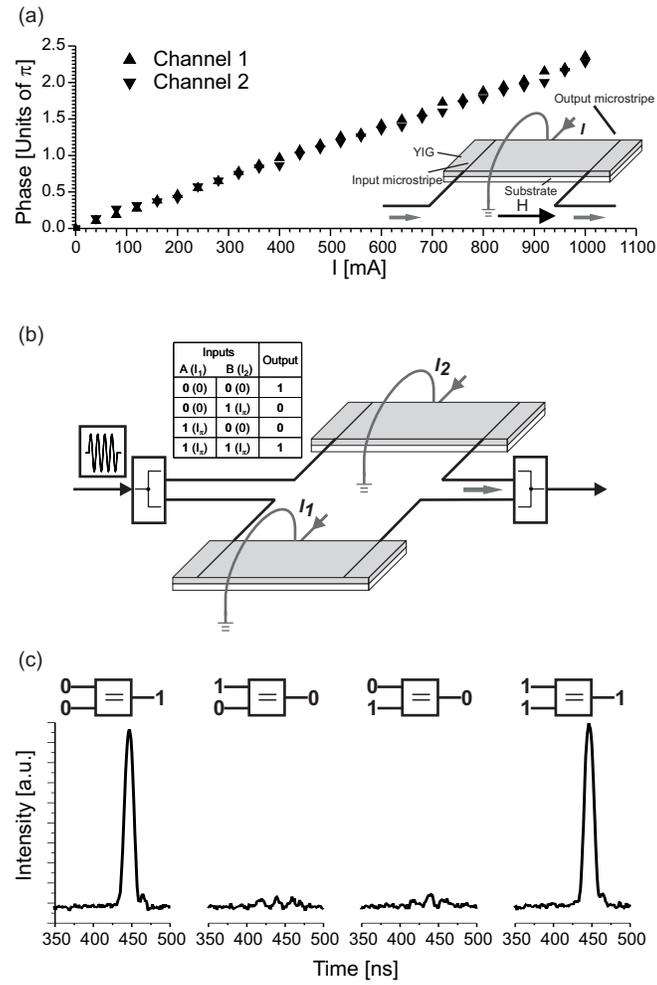}
  \caption{XNOR gate. (a) Inserted phase versus current for the current controlled
  spin wave phase shifters (CPS) used to construct the XNOR gate prototype. It is clearly
  visible that the phase shifts in both arms (channels) are identical. The
  inset shows the phase shifter geometry. (b) Spin-wave XNOR gate geometry.
  The currents $I_{1}$ and $I_{2}$ represent the logical inputs
  ($0\,\mathrm{A}$ corresponds to $\mathbf{0}$, $I_{\pi}$ corresponds
  to $\mathbf{1}$), the spin-wave interference signal represents the logical
  output. Inset: Truth table for an XNOR gate. (c) Gate output signals for input signals shown in the
  diagrams.} \label{XNOR}
\end{figure}

Figure\ \ref{XNOR}(b) shows the principle setup of an exclusive not
OR (XNOR, also called logical equality) gate. It consists of two
arms of a spin-wave Mach-Zehnder interferometer implemented as
ferrite film structures. Spin waves are inserted in both arms using
microstrip antennas connected to a common microwave pulse source,
thus guaranteeing the same phase in both arms. The spin waves are
phase-coherently detected using microstrip antenna detectors. The
signals of both arms are brought to interference electronically. The
phase accumulated by the spin waves on their paths through the two
arms is controlled by applying dc currents $I_1$ and $I_2$ to the
arms. Figure\ \ref{XNOR}(a) shows phase inserted due to a current in
an interferometer arm. One sees a linear dependence of the
accumulated phase on the current. One also sees that the phase
characteristics in both arms are identical.

The currents $I_1$ and $I_2$ serve as \textit{logical} inputs, where
a logical zero is represented by $I=0\ \mathrm{A}$ and a logical one
by the current $I_{\pi}$, necessary to create a phase shift of
$\pi$. The microwave pulses at the \textit{physical} input of the
Mach-Zehnder interferometer represent rate pulses. The amplitude of
the microwave interference pattern at the interferometer physical
output serves as the logic output. Destructive interference (i.e.,
zero intensity) represents a logical zero and constructive
interference (i.e., high intensity) represents a logical one. It is
assumed that both arms are identical and thus both spin waves reach
the output with identical amplitude and phase if no current is
applied. If we now apply a current of strength $I_{\pi}$ to one of
the conductors (i.e., logical one to one input, zero to the other)
one of the spin waves is shifted in phase by $\pi$ which will lead
to destructive interference (logical zero). If the currents in both
conductors are identical (i.e., logical zero or one to both inputs)
the interference will remain constructive (logical one). This
behavior is summarized in the inset to Fig.\ \ref{XNOR}(b). It
resembles a XNOR gate.

The physical mechanism underlying the control of the spin-wave phase
by a dc current through a conductor can be explained as follows. Due
to the Oersted field of the conductor the dispersion curve is
shifted along the frequency axis and thus the carrier wavenumber of
the spin-wave pulse is changed. This results in a change of
spin-wave phase. The accumulated phase is linearly proportional to
the current strength. As shown in Ref.\ \cite{7}, in the case of
wide conductors it is possible to shift the spin wave phase by $\pi$
without introducing a noticeable decrease in the output spin wave
amplitude due to reflection back from the region of induced field
inhomogeneity.

To simplify demonstration of functionality of the XNOR gate we used
an up-scaled prototype, compared to the common sizes of magnetic
elements used in other logic schemes \cite{2,3}. The physical
principles underlying the device performance would remain
practically the same if the prototype is implemented with micrometer
sizes. As a medium for spin wave propagation we use two 6\ \textmu m
thick and 1.5\ mm wide yttrium iron garnet (YIG) spin-wave
waveguides (see Fig.\ \ref{XNOR}). In YIG films spin waves can
travel over several tens of millimeters due to its extremely low
magnetic damping. Conventional Permalloy films would allow for spin
wave propagation over distances up to several tens of micrometers
and are well suited for miniature devices.

Also for the sake of simplicity the physical input and output of the
Mach-Zehnder interferometer are implemented as microwave microstrip
antennas. To transform microwave signals into spin-wave pulses in
the YIG stripes and back into microwave signals microwave
transducers are utilized in the form of 50\ \textmu m wide
microstrip antennas placed 8\ mm apart. (Note that the
interferometer physical input and output can be implemented as
all-spin wave waveguides, as shown in Ref.\ \cite{4}. This will make
them compatible with a spin wave data bus, as recently suggested in
Ref. \cite{8}. In such a structure the microstrip transducers are
not needed.) Wire loops (wire diameter is 0.5\ mm) for spin wave
phase control are placed between the input and the output microwave
transducers. The spin-wave pulses in both interferometer arms are
20\ ns in length. The microwave carrier frequency of the pulses is
7.132\ GHz and the applied bias magnetic field is 1850\ Oe. Control
current pulses of 990\ ns duration are applied to the conductors.
Observe that the length of the current pulses can be reduced to the
length of the spin-wave pulses at the cost of an increased current
necessary to create the $\pi$ shift due to the then shorter
interaction time.

Figure\ \ref{XNOR}(c) demonstrates the microwave signal at the
output of the prototype gate for different input configurations. The
expected behavior is clearly visible.

\begin{figure}[h!tp]
  \centering
  \includegraphics{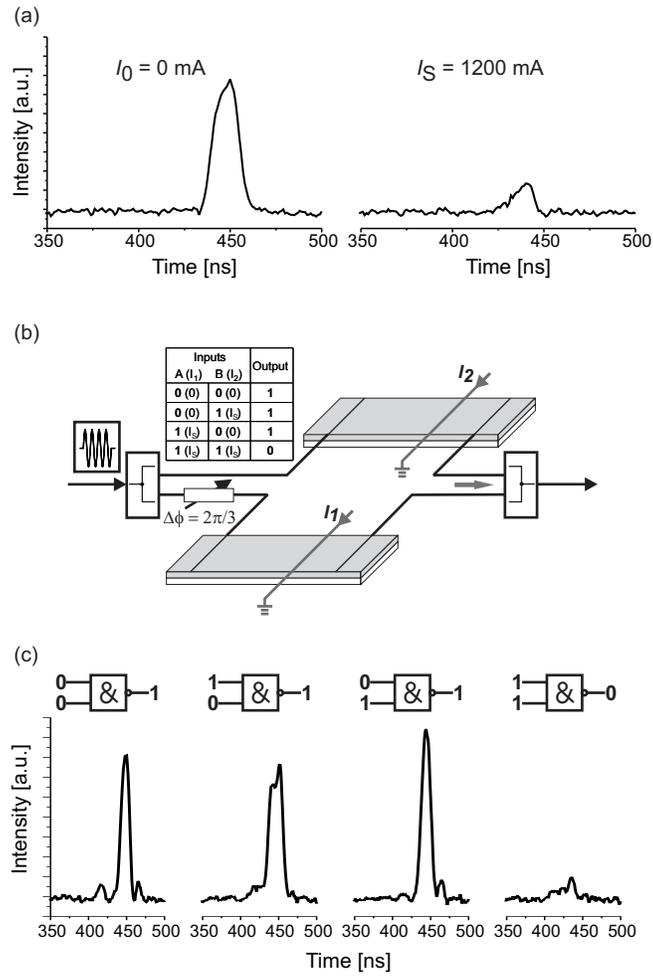}
  \caption{NAND gate. (a) Demonstration of a spin-wave switch. Left part: Output
  signal without applied current. Right part: Output signal with
  applied current. Suppression of the output pulse is clearly
  visible. (b) Geometry of a spin-wave NAND gate. The currents $I_{1}$ and
  $I_{2}$ represent the logical inputs ($0\,\mathrm{A}$ corresponds
  to $\mathbf{0}$, $I_{S}$ corresponds to $\mathbf{1}$); the spin-wave
  interference signal represents the logical output. Inset: Truth table for a
  NAND gate. (c) Gate output signals for
  input signals as shown in the diagrams.}\label{NAND}
\end{figure}

In our previous works \cite{9,10} we show that one can control the
spin wave amplitude by inserting a highly localized magnetic field
inhomogeneity. The inhomogeneity can be created by  a dc current
through a narrow conductor placed on the surface of a ferrite film.
A conductor with a width of 100\ \textmu m produces a highly
localized Oersted field. Independently of the sign of this
additional Oersted field with respect to the applied static field, a
spin-wave pulse can be reflected efficiently from such a strong
inhomogeneity. In this work we choose the current direction which
produces an Oersted field in the direction opposite to the applied
field. By applying a relatively small current it is possible to
shift the dispersion curve so that the carrier frequency of the spin
wave pulse incident onto the inhomogeneity is no longer inside of
the frequency band of spin waves. Spin waves cannot propagate
through this prohibited zone, they can only tunnel. This introduces
a strong back reflection of the incident pulse and one observes a
strong change in the intensity of the spin wave \cite{9}. It is
important to notice that one can reduce the spin wave amplitude to
nearly zero and thus use this setup as a spin-wave switch. The
functionality of such a switch is experimentally demonstrated in
Fig.\ \ref{NAND}(a).

The realization of a logic NOT AND (NAND) gate is demonstrated in
Fig.\ \ref{NAND}(b). The setup mainly consists of a Mach-Zehnder
interferometer, but this time the phase shifters in the arms are
replaced by  switches. Similar to the XNOR gate the logical output
is implemented by the interference signal, while the inputs are
implemented by the currents. Logical zero is represented by
$I_{0}=0\,\mathrm{A}$, and logical one by the current $I_{S}$,
necessary to suppress the spin wave pulse transmission.

Experimentally measured output interferometer pulses are shown in
Figure\ \ref{NAND}(c). If a current is applied to only one of the
switches (logical one to one input, zero to the other) a microwave
pulse of a large intensity is transmitted (logical one at the
output). The same takes place if no current is applied to both arms
(logical zero to both inputs). To ensure that the intensity of the
output microwave pulse is the same in this case as in the two other
cases of a logical one at the output an additional permanent phase
shift of $2\pi/3$ is introduced in one of the interferometer arms.
In this prototype we use an external coaxial microwave phase
shifter. However a current controlled \textit{spin wave} phase
shifter similar to the one implemented in the XNOR gate prototype
(Fig.\ \ref{XNOR}(b)) may be easily integrated into the ferrite film
structure instead. Any other possibility to create an additional
phase shift (e.g., make the spin-wave propagation paths in the arms
differ by $2\pi/3$ by slightly increasing the length of one of the
arms or using a thin permanent magnet to apply a small additional
bias magnetic field to one of the arms) is also possible. A current
applied to both switches (logical one to both inputs) leads to a
nearly complete suppression of the output signal (logical zero). The
described behavior (summarized in the inset of Fig.\ \ref{NAND}(b))
is the one of a NAND gate.

In conclusion, we experimentally demonstrated the functionality of
universal NAND and a XNOR logical gate using spin-waves propagating
in a magnetic film interferometer based on YIG. By changing the used
spin-wave waveguides (e.g., to Permalloy) a downscaling of the
presented devices down to micrometer-scale dimensions should be
possible and would open up a new approach to logic structures on the
micrometer scale.

\begin{acknowledgments}
Support by the Deutsche Forschungsgemeinschaft (Graduiertenkolleg
792), the Australian Research Council, and the European Community
within the EU-project MAGLOG (FP6-510993) is gratefully
acknowledged.

The work and results reported in this publication were obtained with
research funding from the European Community under the Sixth
Framework Programme Contract Number 510993: MAGLOG. The views
expressed are solely those of the authors, and the other Contractors
and/or the European Community cannot be held liable for any use that
may be made of the information contained herein.
\end{acknowledgments}

\bibliography{}

\newpage

\end{document}